\documentclass{elsart5p}
\usepackage[dvips]{graphicx}
\usepackage{amssymb}
\usepackage{subfigure}

\begin{document}

\begin{frontmatter}
\title{Measuring energy dependent polarization in soft
gamma-rays using Compton scattering in PoGOLite}
\author[su]{M.~Axelsson\corauthref{cor}}, \author[kth,su]{O.~Engdeg\aa rd}, \author[kth,su]{F.~Ryde}, \author[su]{S.~Larsson}, \author[kth]{M.~Pearce}, \author[hel,su]{L.~Hjalmarsdotter}, \author[kth]{M.~Kiss}, \author[kth]{C.~Marini~Bettolo}, 
\author[tit]{M.~Arimoto},
\author[su]{C.-I.~Bj\"ornsson},
\author[kth]{P.~Carlson},
\author[hiro]{Y.~Fukazawa},
\author[slac,kav]{T.~Kamae},
\author[tit]{Y.~Kanai},
\author[tit]{J.~Kataoka},
\author[tit]{N.~Kawai},
\author[kth]{W.~Klamra},
\author[slac,kav]{G.~Madejski},
\author[hiro]{T.~Mizuno},
\author[slac]{J.~Ng},
\author[slac,kav]{H.~Tajima},
\author[isas]{T.~Takahashi},
\author[hiro]{T.~Tanaka},
\author[tit]{M.~Ueno},
\author[haw]{G.~Varner},
\author[hiro]{K.~Yamamoto}

\corauth[cor]{Corresponding author.\\ \textit{E-mail:}~magnusa@astro.su.se}
\address[su]{Stockholm Observatory, AlbaNova, SE-106 91 Stockholm, Sweden}
\address[kth]{Physics Department, Royal Institute of Technology, AlbaNova, SE-106 91 Stockholm, Sweden}
\address[hel]{Observatory, PO Box 14, FIN-00014 University of Helsinki, Finland}
\address[tit]{Tokyo Institute of Technology, 2-12-1 Ookayama, Meguro, Tokyo 152-8551, Japan}
\address[hiro]{Hiroshima University, Physics Department, Higashi-Hiroshima 739-8526, Japan}
\address[slac]{Stanford Linear Accelerator Center, 2575 Sand Hill Road, Menlo Park, CA 94025, USA}
\address[kav]{Kavli Institute for Particle Astrophysics and Cosmology, Stanford University, Stanford, CA 94305, USA}
\address[isas]{Institute of Space and Astronautical Science, Japan Aerospace Exploration Agency, Sagamihara 229-8510, Japan}
\address[haw]{Department of Physics and Astronomy, University of Hawaii, 2505 Correa Road, Honolulu, HI 96822, USA}

\begin{abstract}
Linear polarization in X- and gamma-rays is an important diagnostic of many astrophysical sources, foremost giving information about their geometry, magnetic fields, and radiation mechanisms. However, very few X-ray polarization measurements have been made, and then only mono-energetic detections, whilst several objects are assumed to have energy dependent polarization signatures.

In this paper we investigate whether detection of energy dependent polarization from cosmic sources is possible using the Compton technique, in particular with the proposed PoGOLite balloon-experiment, in the 25--100 keV range. We use Geant4 simulations of a PoGOLite model and input photon spectra based on Cygnus X-1 and accreting magnetic pulsars (100\,mCrab). Effective observing times of 6 and 35 hours were simulated, corresponding to a standard and a long duration flight respectively. Both smooth and sharp energy variations of the polarization are investigated and compared to constant polarization signals using chi-square statistics.

We can reject constant polarization, with energy, for the Cygnus X-1 spectrum (in the hard state), if the reflected component is assumed to be completely polarized, whereas the distinction cannot be made for weaker polarization. For the accreting pulsar, constant polarization can be rejected in the case of polarization in a narrow energy band with at least 50\% polarization, and similarly for a negative step distribution from 30\% to 0\% polarization.

\end{abstract}

\end{frontmatter}

\section{Introduction}
\label{intro}

In the areas of spectral and temporal studies, X-ray and gamma-ray astronomers have 
been given a wealth of data on a wide range of objects. Polarization has 
long been predicted to play a crucial role in 
determining physical and geometrical parameters in many astrophysical
sources, thereby discriminating among current models. However, there have 
so far been very few measurements of polarization at these energies. In light of this, the
possibility to detect energy dependent polarization has hardly been 
discussed at all in the literature. In this paper, we present the results 
from simulations of a dedicated soft gamma-ray polarimeter using Compton 
scattering, and 
study the response when the degree of polarization varies with the energy 
of the emitted photons. While energy-dependent polarization is expected 
from many sources, its detection requires an instrument of sufficiently 
good energy response. The Compton polarimeter presented in this paper 
utilizes plastic scintillators, which are relatively inefficient for energy 
depositions below a few keV.
Thus, simulations are necessary to determine how sensitive the instrument 
is and how large variations must be for detection. 

We begin by describing the organisation of the paper. The remainder of this
section is devoted to giving a background of polarimetry in the X/$\gamma$-ray
regime, and an overview of the scientific motivation for such measurements.
In Section~\ref{compton} we focus on polarimetry using Compton scattering
and describe an instrument design based on this technique. We then present
the set-up of our simulation of the instrument in Section~\ref{simulations}, 
and the results of the simulations in Section~\ref{results}. Finally, in 
Sections~\ref{discussion} and \ref{conclusions}, we discuss and summarise 
our results.

\subsection{Measurement of polarization}

The aim of any polarimetric measurement is to determine the degree and
direction of polarization of incident radiation. When combined with the
traditionally measured quantities of energy and time, polarimetry has
the potential to double the parameter space available. As such, it can
be a powerful tool to discriminate between physical models proposed for
a given source.

Historically, polarimetry has proven very successful at optical 
and radio wavelengths. In these bands, it has been 
extensively used to probe both radiation physics and geometry of
sources (see, e.g., \cite{tin05}). In the X-ray regime, however, the 
results are more meagre. Early rocket observations measured X-ray
polarization from the Crab Nebula \cite{nov72}. This result was
later confirmed by the Orbiting Solar Observatory 8 (OSO-8, measuring 
a polarization degree of $19.2 \% \pm 1.0 \%$, \cite{weis76,weis78}), 
the only satellite 
mission carrying a dedicated polarimeter to date. As the design
was based on Bragg reflection on graphite crystals, the energies
probed were constrained to 2.6 keV and 5.2 keV.

A number of new polarimetric instruments, designed to work in the X/$\gamma$-ray regime, have recently been proposed. 
These include POLAR (10--300 keV, \cite{pro05}), GRAPE (50--300 keV, \cite{leg05}), PHENEX (40--300 keV, \cite{gun03}),  CIPHER (10\,keV -- 1\,MeV, \cite{sil03}), and POLARIX (1.5--10 keV, \cite{cos06}). In this paper we present PoGOLite, a Compton polarimeter currently under construction \cite{kam07}.

\subsection{Expected objects of interest}
\label{objects}

The lack of polarimetric measurements in X-rays is not due to a lack
of potential targets. Indeed, from a theoretical point of view there are 
many sources that are expected to display detectable degrees of 
polarization. Over the past decades, there have been publications 
discussing the potential for polarization in sources such as X-ray
binary (XRB) systems, active galactic nuclei (AGN), accretion and 
rotation powered pulsars as well as cataclysmic variables (CVs); 
see e.g., \cite{rees75,mesz88,agol96,matt04,viir04}. Other work has 
focused on the processes producing polarized radiation, either 
the radiative processes themselves (e.g., synchrotron and 
non-thermal bremsstrahlung, \cite{wes59,gne74}), 
or processes such as reflection/asymmetric scattering (e.g., 
\cite{sun85,matt93,ogur00}), strong-field gravity \cite{conn80,kar05} and 
vacuum birefringence in strong magnetic fields \cite{nov77}.

In most sources, polarization is not expected to remain constant 
with energy. 
An example is radiation from strongly magnetized plasmas
where the polarization may change dramatically near the cyclotron
resonance energy.
It is therefore important to understand not only what degree of polarization
is needed for detection, but also how sensitive a given instrument will be to 
the changes of polarization with energy. To study such effects we have chosen 
to simulate two example sources: Cygnus X-1 and an accreting magnetic 
neutron star.

\subsubsection{Cygnus X-1}

\begin{figure}
\includegraphics[width=9cm]{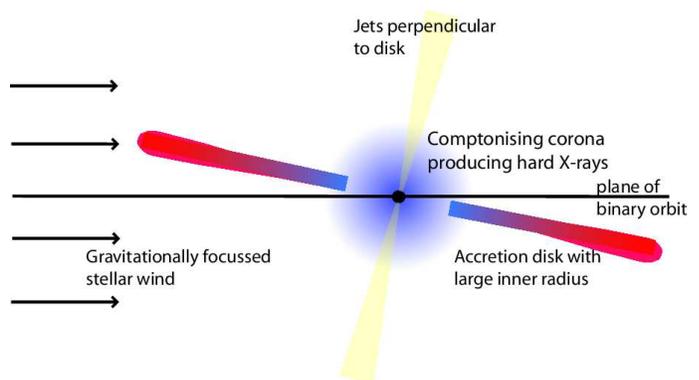}
\caption{Likely geometry in the hard state of Cyg~X-1. Mass being accreted
forms an accretion disc around the compact object. In the inner regions,
there is a hot inner flow/corona. Soft seed photons from the disc may 
be Comptonized in the hot flow. A fraction of the resulting hard photons can 
then be reflected off the disc, giving a net polarization.}
\label{cygx1geo}
\end{figure}

Cygnus~X-1 is a high-mass XRB where the compact object is believed to be 
a black hole.
The source exhibits two main spectral states, commonly referred to as hard
and soft. Most of the time is spent in the hard state. Several models 
have been proposed to explain the observed states and transitions.
The two main components of such models are usually a geometrically thin, 
optically thick accretion disc and a hot inner flow or corona \cite{pou98}.
A schematic picture of a likely geometry in the hard state is shown in 
Fig.~\ref{cygx1geo}.
\noindent Soft X-rays are produced in the accretion disc, and may then be 
Comptonized in the hot inner flow/corona. A fraction of the hard radiation 
can be reflected off the accretion disc before reaching the observer.
Polarization from this system may arise through several processes. 
In this paper, we will focus on the polarization introduced
by the reflection (for more details, see, e.g., 
\cite{matt93,matt93b,bal01}). In Cygnus~X-1, this contribution
is strongest in the energy range of $\sim$\,20--100\,keV. The polarization 
degree is expected to vary with energy, following the
relative strength of the reflection component.

\subsubsection{Accreting magnetic neutron stars}
In many high-mass XRBs the accreting object is a highly magnetic
neutron star. The strong magnetic field, \mbox{$\sim10^{13}$ gauss} at the surface, 
directs the accretion flow towards the magnetic poles of the star. Most 
of the accretion energy is released just above the polar cap
where the emission and propagation of radiation is directly connected
to the magnetic field as well as the local properties of the plasma. For 
a number of sources cyclotron spectral features have been observed in 
hard X-rays, and from these, magnetic field strengths have been deduced. 
The X-rays are expected to be polarized and the degree, angle and energy 
dependence of the polarization will depend on the physical
conditions in the emission region \cite{and04}. Measurements of the detailed
polarization properties would therefore provide a new and very powerful 
probe of the radiating plasma near the surface of the neutron star.

\section{The Compton technique}
\label{compton}

Apart from the special case of Bragg reflection, all three main physical 
processes of photon-matter interaction in the X/$\gamma$-ray regime may
be used in polarimetry: photoabsorption, Compton scattering and pair production. Each of these 
preserves information on the polarization of the incoming radiation. 
For photon energies between $\sim 100$ keV and 1 MeV, Compton scattering
is the dominant process. In this section, we will briefly outline the 
theoretical basis for a polarimeter based on Compton scattering, and present a design
for a dedicated polarimeter based on this technique.

\subsection{Basic principle}

The differential cross section for Compton scattering is given by:
\begin{equation}
\frac{\textrm{d}{\sigma_{\rm cs}}}{\textrm{d}{\Omega}}=\frac{1}{2} r_e^2 \frac{E^2}{E_0^2} \left[ \frac{E}{E_0}+\frac{E_0}{E}-2\sin^2\theta \cos^2\phi\right] \: ,
\end{equation}
\noindent
where $r_e$ is the classical electron radius, $E_0$ and $E$ are the photon frequency before and after scattering,
$\theta$ is the angle between incident and scattered direction, and $\phi$
is the azimuthal scattering angle relative to the plane of polarization. When
projected on a plane, the angle of scattering will thus be modulated as
$\cos^2\phi$.

To measure the scattering angles, it is necessary to detect both the 
site of scattering and that of photoabsorption. If more than two scattering 
sites are identified, the relative energy depositions can be used to help distinguish between Compton scattering and photoelectric absorption sites. Some form of segmentation of the detector is necessary to provide spatial resolution, required to determine the positions of the signals.

\subsection{PoGOLite}
\label{pogo}

\begin{figure}
\begin{center}
\includegraphics[width=8cm]{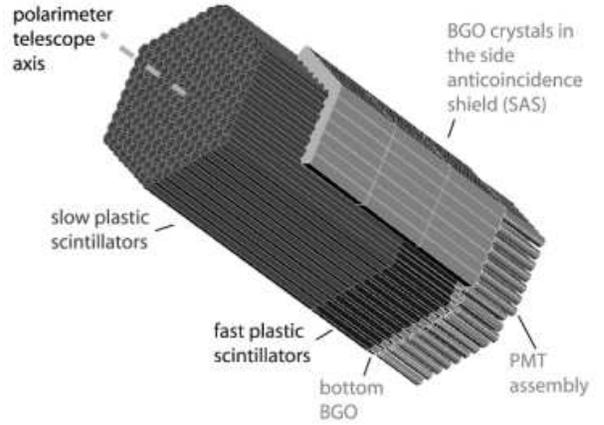}
\end{center}
\caption{The design of the PoGOLite instrument. The side anticoincidence
shield has been partially cut away for clarity. The total length of the instrument will be $\sim 100$\,cm.}
\label{pogo1}
\end{figure}

The Polarized Gamma-ray Observer - Light weight version (PoGOLite) is a 
balloon-borne
polarimeter, planned for launch with a stratospheric balloon in 2009. Figure~\ref{pogo1} shows the design
of the instrument.
\noindent
The instrument consists of 217 phoswich 
detector cells (PDCs) arranged in a hexagonal pattern. Each PDC is made up 
of a hollow slow
scintillator tube, a fast scintillator detector, a bottom bismuth germanate
(BGO) crystal, and a photomultiplier tube (PMT). Signals from the different 
optical components are distinguished using a pulse shape discrimination 
technique based on
the different scintillation decay times of the materials \cite{kan07}. The 
configuration is surrounded by an anticoincidence shield made of BGO 
crystals. Together with the bottom BGO crystals, this allows side and back 
entering photons and cosmic rays to be rejected.

The hollow slow scintillator tube acts as an active collimator. Photons or 
charged particles entering the instrument off-axis will be registered in 
the slow
scintillator and can be rejected. The desired events are from photons that 
enter cleanly through the slow scintillator and scatter in the 
fast scintillator. After scattering, the photon may be absorbed in one of 
the neighbouring fast scintillator cells, allowing the azimuthal scattering 
angle to be determined. 

The well-type design of PoGOLite allows for efficient background rejection 
\cite{kam96,tak92}, 
and gives a field of view of \mbox{$\sim5\;{\rm deg}^2$}. This allows the instrument to be 
accurately pointed at specific sources. As both the initial Compton 
scattering and subsequent photon absorption occur in the same material 
(the plastic fast scintillator), the effective energy range is determined 
by the cross-sections for both these processes, as well as
the background. PoGOLite will have an energy range of $\sim$\,25--100\,keV, 
which is lower than the range where Compton scattering dominates. A more 
detailed description of the instrument may be found in \cite{kan07,lp04}.

The capability of PoGOLite to measure the energy dependence of
polarization is limited both by the signal-to-background ratio and the 
energy resolution. Due to redistribution, some of the higher energy photons 
will
produce events at lower energies. The flux and polarization in the low 
energy band will therefore be affected by the spectrum at higher energies 
but not vice-versa. The energy response has been carefully simulated using
Geant4 \cite{olle}.

\begin{figure}
\includegraphics[width=8cm]{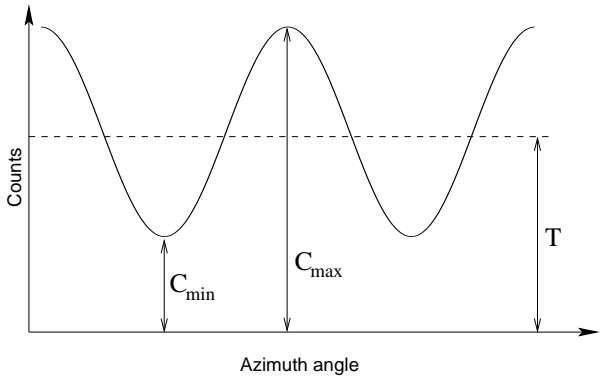}
\caption{Simplified sketch of the distribution of scattering angles, used to
determine the modulation factor. The maximum ($C_{\rm max}$), minimum ($C_{\rm min}$) and average ($T=[C_{\rm max}+ C_{\rm min}]$/2) values of the distribution are indicated.}
\label{modsketch}
\end{figure}

Figure~\ref{modsketch} shows a hypothetical distribution of azimuthal scattering angles. The maximum ($C_{\rm max}$) and minimum ($C_{\rm min}$) values of the distribution and the average ($T=[C_{\rm max} + C_{\rm min}]$/2) can be used to define a modulation factor: 

\begin{equation}
M \equiv \frac{C_{\rm max}-C_{\rm min}}{C_{\rm max}+C_{\rm min}}=\frac{C_{\rm max}-C_{\rm min}}{2T}\;.
\label{modfac} 
\end{equation} 

\noindent The modulation factor is determined by fitting the following function to the distribution of azimuthal scattering angles  

\begin{equation}
f(x)=T(1+M\cos(2x+2\alpha))\: ,
\label{distfunction}
\end{equation}
with angle $x$ (a function variable, not a fitting parameter), average $T$, modulation factor $M$, and polarization angle $\alpha$. 

In this work, the modulation factor is the discriminator between different polarization models. If the response of the instrument to a 100\% polarized source is known, the modulation factor can be used to determine the polarization of the incoming photon beam~\cite{lei97}.

\section{Simulations}
\label{simulations}
In this section we will describe the setup of our simulations. The source models used as input are also presented, as well as the background considered.

\subsection{Geant4}
Geant4\footnote{http://geant4.cern.ch} is a multi-purpose software package for simulating particles travelling through and interacting with matter, using Monte Carlo techniques \cite{geant4}. The standard Geant4 package was earlier found \cite{miz05} to have incorrect implementations concerning photon polarization in Compton and Rayleigh scattering; the Geant4 version used here is a corrected version of 4.8.0.p01.

\subsection{Simulation setup}
The Geant4 implementation includes the essential parts of PoGOLite: 217 PDCs with slow and fast plastic scintillators and bottom BGO crystals together with a BGO side shield. The model has no PMTs, and uses a solid BGO side shield instead of discrete pentagonal bars (cf. Fig.~\ref{pogo1}). Layers of tin (50$\,\mu$m) and lead (50$\,\mu$m) surrounding each slow scintillator and the ${\rm BaSO_4}$ coating (200$\,\mu$m) of the BGO crystals are included. The mechanical support structure is not represented. 

During the simulation, separate photons are generated with random energies from a spectral model. An event is triggered by a hit in two or three of the fast plastic scintillators. The following is saved as output data for each event: information about the original gamma momentum, the ID-number of the cells that had an interaction (ranging from 1 to 217) and the energy deposited in each cell. These data are preprocessed to simulate the resolution of the PMTs, as described in Sect.~\ref{data}. 

\subsection{Source Models}
As stated in Sect.~\ref{objects}, two sources were considered: Cygnus~X-1 (in 
the hard state) and an accreting neutron star. Below we describe the model
used for the incident radiation and polarization in each case. As shown in
\cite{kam07}, PoGOLite is expected to detect polarization in both 
these sources; what we are investigating is the sensitivity to changes in 
polarization degree with energy. 

\subsubsection{Cygnus X-1}
\begin{figure}
\includegraphics[width=9cm]{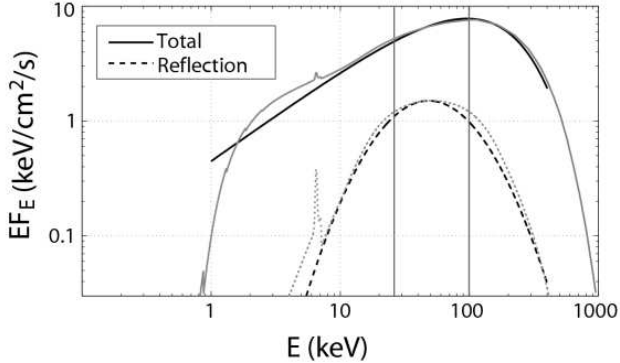}
\caption{Observed radiation spectrum and input model used in the simulations of observations of Cygnus~X-1. \textit{Gray lines:} Typical radiation spectrum of Cygnus~X-1 in the hard state. \textit{Black lines:} Assumed input spectrum: a cut-off power law with index $\alpha = -1.2$ and cut-off $E_{\rm cut}=120$ keV, normalized to match the measured flux. The reflection component is shown, and the energy range of PoGOLite is indicated by vertical lines.}
\label{cygx1model}
\end{figure}

For our simulations of Cygnus X-1 we used an input spectrum of a power-law, with photon index $\alpha=-1.2$ and an exponential cutoff at energy $E_{\rm cut}=120$ keV. It was normalised to match the observed spectrum of Cygnus~X-1. The spectrum of the reflection was approximated by the logarithmic quadratic curve
\begin{equation}
EF_E=10^{ -c(\log E-\log a)(\log E-\log b) },
\end{equation}
with $a=24$, $b=98$ and $c=1.89$. Figure~\ref{cygx1model} shows the observed
radiation of Cygnus~X-1, and our model of the total spectrum as
well as that assumed for the reflection component. 

In our simulations, the polarization is assumed to arise due to the reflection component. Two scenarios were tested: 100\% and 20\% polarization for the reflection component, with unpolarized direct emission. This corresponds to a total average polarization around 17\% and 3\% respectively. The energy dependent polarization $\Pi(E)$ used as input was set to the relative strength of the reflection component compared to the total flux, scaled down in the case of 20\% polarization. Simulations were performed for effective observing times of 6 hours and 35 hours. These times are chosen as realistic estimates for short and long duration balloon flights, respectively.

\subsubsection{Accreting Magnetic Neutron Star}

In the case of the neutron star, we study the observability of energy dependent
effects by simulations of three different idealized polarized spectra:
\begin{itemize}
\item Polarization in a narrow band.
\item Polarization only at low energies. 
\item Polarization only at high energies.
\end{itemize}
The neutron star spectrum was in all cases approximated with an exponentially cut-off 
power law, with index \mbox{$\alpha=-1.1$} and energy cut-off at $E_{\rm cut}=70$\,keV. It was 
normalized to correspond to a 100\,mCrab source.

Assuming a cyclotron energy $E_c$ at 50 keV, we use three toy models of the polarization energy dependence $\Pi(E)$, with $\Pi_{\rm max}\equiv p\,\%$:
\begin{itemize}
\item A Gaussian peak centred at 50 keV, $G_p$, modelling a rise in polarization from 0\% to maximum $p\,$\%, using the Gaussian curve
  \begin{equation}
    \Pi(E)=pe^{\frac{-(E-50)^2}{2\sigma^2}}\,\%
\label{eq:Gp}
  \end{equation}
  with $E$ measured in keV and $\sigma=5$ keV.
\item Two step functions, $S_p$ and $S_{-p}$, with polarization
  \begin{equation}
    \Pi=\left\{ \begin{array}{ll}
	0\% & \textrm{if $E<50$ keV} \\
	p\,\% & \textrm{if $E\ge50$ keV}
      \end{array} \right.
  \end{equation}
for $S_p$, and
\begin{equation}
    \Pi=\left\{ \begin{array}{ll}
	p\% & \textrm{if $E<50$ keV} \\
	0\,\% & \textrm{if $E\ge50$ keV}
      \end{array} \right.
  \end{equation}
for $S_{-p}$. Simply put, $p$ is the jump in polarization that occurs at $E=50$ keV.
\end{itemize}

\begin{figure}
\includegraphics[width=8cm]{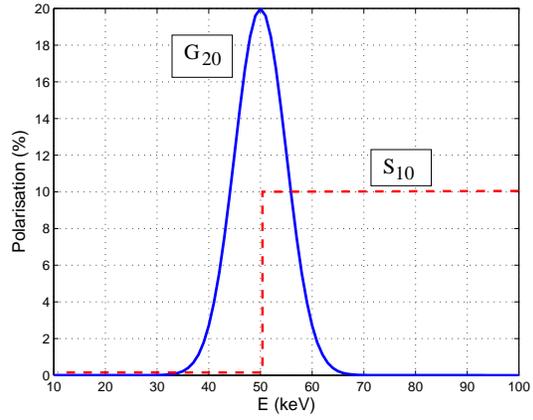}
\caption{Example of assumed energy dependence of the polarization fraction in the case of an accreting magnetic neutron star. The figure shows a Gaussian curve with $\Pi_{\rm max}=20\%$ (G$_{20}$), and a positive step with $\Pi_{\rm max}=10\%$ (S$_{10}$).}
\label{nsmodels}
\end{figure}

Figure~\ref{nsmodels} illustrates examples of $G_{20}$ and $S_{10}$, the Gaussian and positive steps with maxima 20\% and 10\% respectively. In the simulations, the values $p=\{10, 20,30,40,50\}$\% were used, each assuming an observation time of 35 hours.

\subsection{Background}

Balloon-borne gamma-ray polarimetry measurements are subject to several significant sources
of background. Through the use of the well-type phoswich detector technique, the PoGOLite 
instrument has been designed to reduce these backgrounds, allowing 10\% polarization of a 
100~mCrab source to be measured in one 6~hour balloon observation in the 25--100\,keV energy 
range. The basic phoswhich design was used in the WELCOME series of balloon-borne 
observations and allowed effective background suppression \cite{kam92,kam93,tak93,gun92,gun94,miy96,yam97}. The concept was subsequently improved and effectively used in a satellite instrument, the Suzaku Hard 
X-ray Detector (HXD) \cite{kam96,mak01,kok04,suz05}.

The background to PoGOLite measurements can arise from charged cosmic rays, neutrons 
(atmospheric and instrumental) and gamma-rays (primary and atmospheric).  The background 
from charged cosmic rays (predominantly protons, $\sim$90\%, and helium nuclei, $\sim$10\%) 
is rejected by the BGO anticoincidence shields and slow plastic collimators. 
Cosmic rays are minimum ionizing particles and can be identified through their relatively 
large energy deposits. The background presented by atmospheric neutrons and neutrons 
produced in the PoGOLite instrument and surrounding structures is currently being studied 
in detail \cite{kaz07}.
For the purposes of the study presented in this paper, particular attention has been paid 
to what is expected to be the dominant background: primary and atmospheric gamma-rays. The 
gamma-ray background rate is estimated from a model derived from measurements taken in Texas with the GLAST 
Balloon Flight Engineering Model~\cite{miz04}.  

The primary gamma-ray component originates outside the atmosphere, i.e., above PoGOLite. The angular distribution of the radiation is uniform within the hemisphere above PoGOLite. The energy spectrum is 
modeled by a doubly-broken power-law with breaks at 50~keV and 1~MeV~\cite{sre98}. 

Secondary gamma-rays are created in the Earth's atmosphere through bremsstrahlung 
interactions of charged cosmic-rays. Two separate components are considered, one directed 
upwards and one downwards. The upward flux is dependent on the zenith angle~\cite{sch77}, 
and the energy spectrum consists of a doubly-broken power-law with breaks at 10 MeV and 1 GeV, and a 
511 keV line from electron-positron annihilation. The downward component is similar, but 
with breaks at 1 MeV and 1 GeV. Energies up to 100 GeV were generated for all components. 
These models are based on data from satellite- and balloon-borne instruments (\cite{geh85} and \cite{sch80}, and references therein).

\begin{figure}
\includegraphics[width=8.5cm]{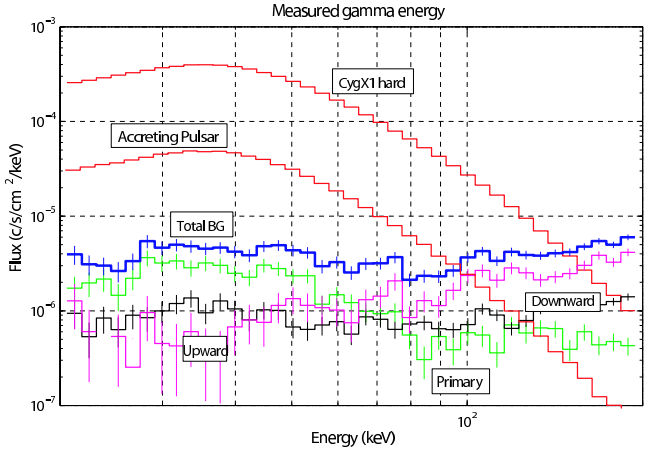}
\caption{Estimated contribution to the gamma-ray background in the PoGOLite energy range ($\sim$\,25--100\,keV). The radiation from the two considered sources, Cygnus~X-1 and an accreting neutron star, are also shown.}
\label{bkgflux}
\end{figure}

Figure~\ref{bkgflux} shows the estimated gamma-ray backgrounds 
compared to the accreting pulsar and Cygnus X-1 models. The total gamma-ray background is 
at the 10 mCrab level.

\section{Analysis and Results}
\label{results}
\subsection{Data Processing}
\label{data}
In the first data processing step, the resolution of the scintillator-PMT assembly is simulated by fluctuating the number of photo-electrons generated in the scintillating materials. It is assumed that when the energy $E$ is deposited in a cell, the average number of photo-electrons generated is $En$, with $n$ set to 0.5 photo-electrons per keV. Now, we fluctuate $En$ to $(En)_{\rm fluct}$ by applying a Gaussian spread. If $En\le10$, we do it in two steps: First we subject it to a Poissonian spread, and thereafter a Gaussian spread with variance $En\sigma^2$, with $\sigma$ set to 0.4. If $En>10$, only a Gaussian spread with variance $En$ is used. If $(En)_{\rm fluct}<0$, then it is set to 0. Finally, we take
\begin{equation}
  E_{\rm mes}=\frac{(En)_{\rm fluct}}{n}
\end{equation}
as the energy actually measured by the PMT in the cell of interest. We reject all fast scintillator interactions with $E_{\rm mes}$ below a certain measurement threshold (2 keV). For the analysis described in this paper, only events with two or three hits in the fast scintillators are retained (the veto logic is not considered at this stage). At PoGOLite energies, more than 80\% of the events are from photons interacting in no more than three detector cells \cite{kis07}.

For two-site events, the chronological order of the two cells does not matter for angle calculation, as the distribution is periodic over the angle $\pi$. In the case
of three hits, we calculate the scattering angle by ignoring the hit with the lowest energy measured, assuming that a low-energy interaction does not affect 
direction much, and derive an angle from the positions of the 
two cells with highest energy deposits. 

Most photons do not scatter very far; about half will only go from one cell to its neighbour. As the range of possible scattering angles resulting in detection in a given adjacent cell is large, this causes strong peaks in each of the six directions corresponding to the neighbouring cells. The PoGOLite instrument will rotate about its axis, causing the range for a given cell to smoothly vary and thereby creating a continuous distribution over angles. In the simulations, the uncertainty of the angle determination is instead approximated by introducing a Gaussian spread to the measured scattering angles. 

In the last steps of data processing we take into account the mass of air in the atmosphere above the balloon, filtering out roughly half of our incident source radiation, assuming the atmospheric overburden 4 g/cm$^2$ at 40 km altitude. We also apply the veto logic, rejecting all events with detection in any slow plastic scintillator or BGO crystal.

\subsection{A $\chi^2$ measure}
To measure the polarization energy dependence, one cannot simply calculate the polarization at certain energies and construct $\Pi(E)$, since the photon energy $E_\gamma$ always will be unknown due to the response of the instrument. Instead we calculate the modulation factor at different \emph{measured} energies, obtaining a curve $M(E_{\rm mes})$. This curve can then be compared with theoretic curves resulting from other models, possibly from the same family of curves, enabling us to reject complete families of energy dependencies. One such family, which we will be concerned with here, is the set of \emph{constant} polarizations.

\begin{figure}
\includegraphics[width=8.5cm]{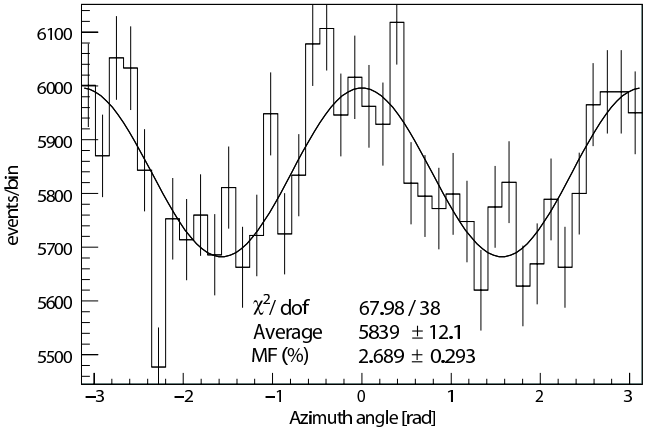}
\caption{Example of the measured distribution of events as a function of
scattering angle (histogram). In this example, the simulated results of a six hour observation of Cygnus~X-1 are shown, assuming a completely polarized reflection component. The data are from the measured energy range of 30--35\,keV. A sinusoidal function is fit to the data (solid
line, cf. Eq.~\ref{distfunction}) and a modulation factor is calculated.}
\label{exfit}
\end{figure}

For a given source model (polarization energy dependence) A, the modulation factor $M$ (Eq.~\ref{modfac}) was fitted at different measured energies, yielding a curve $M_A=M_A(E_{\rm mes})$. Figure~\ref{exfit} shows an example of a modulation curve in the 30--35\,keV band, generated for a six hour observation of Cygnus~X-1 in the case of a completely polarized reflection component. The resulting modulation factor is $2.69 \pm 0.30$. The modulation factor is in this way calculated for each energy band. The result is a curve showing how the modulation factor varies with energy, which can then be compared to the corresponding curves for various models of polarization. 

To test the results against constant polarization, we also generated curves $M_\Pi$ with constant levels of polarization $\Pi$. Since these latter curves should not be thought of as measured, but fluctuation-free theoretical constructs, they were generated by much longer simulations than the observational curves.

A measure of how much two curves differ is defined as
\begin{equation}
  \chi^2_\Pi = \sum_{E_{\rm mes}} \frac{ (M_{A,E_{\rm mes}} - M_{\Pi, E_{\rm mes}})^2}{\sigma_{E_{\rm mes}}^2},
  \label{eq:chi2}
\end{equation}
with $\sigma_i$ as the sum of the two errors in fitting $M_{A,E_{\rm mes}}$ and $M_{\Pi,E_{\rm mes}}$. When this is calculated for all reasonable values of $\Pi$, we can reject the hypothesis of constant polarization if the minimum of $\chi^2_\Pi$ is high enough. For 16 degrees of freedom, corresponding to data points up to 100 keV, the 95\% certainty level requires $\chi^2>26.3$.

\subsection{Cygnus X-1}
\begin{figure*}
  \centering
  \includegraphics[width=8cm]{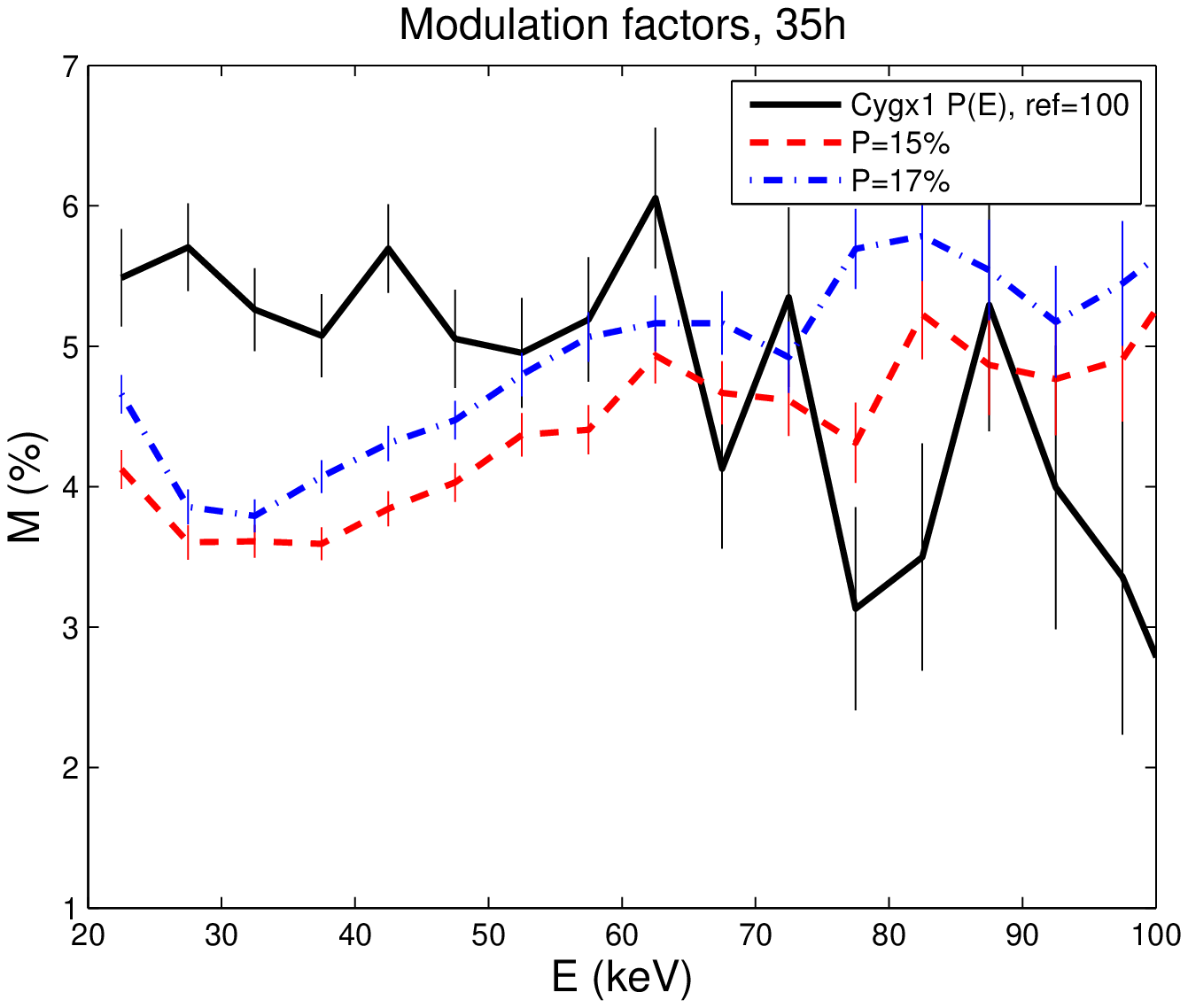}
  \includegraphics[width=7cm]{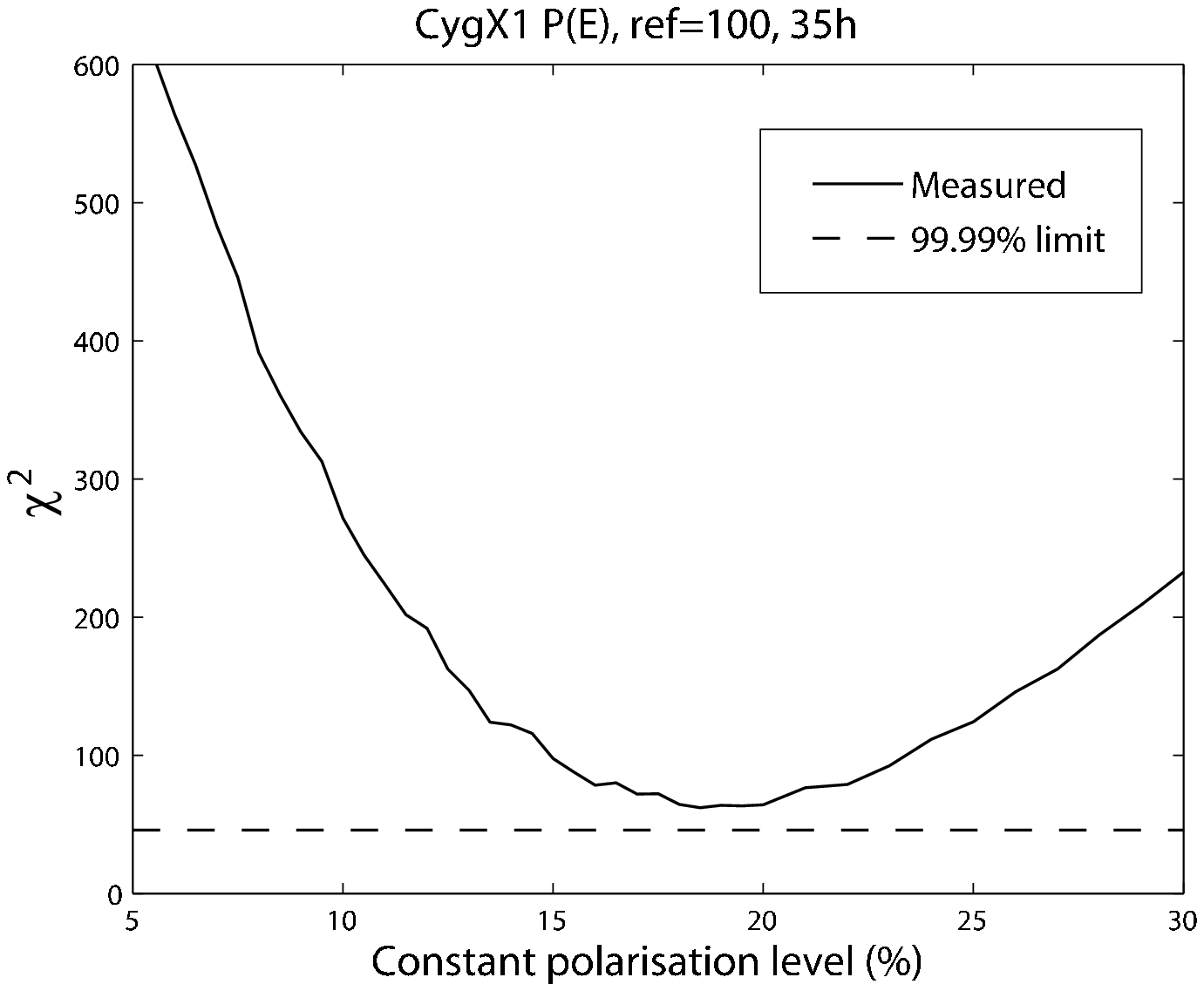}
  \caption{Results from simulations of a 35h observation of Cygnus X-1, with the reflection component assumed to be 100\% polarized. \textit{Left panel:} Expected modulation factor $M(E)$, together with models of constant polarization at 15\% and 17\%. The modulation is fitted in intervals of 5\,keV. \textit{Right panel:} $\chi^2$ values when comparing $M(E)$ for the simulation with different degrees of constant polarization. The dashed line marks the limit for 99.9\% significance in rejecting constant polarization. Constant polarization is rejected with high significance.}
  \label{cyg100results}
\end{figure*}

\begin{figure*}
  \centering
  \includegraphics[width=8cm]{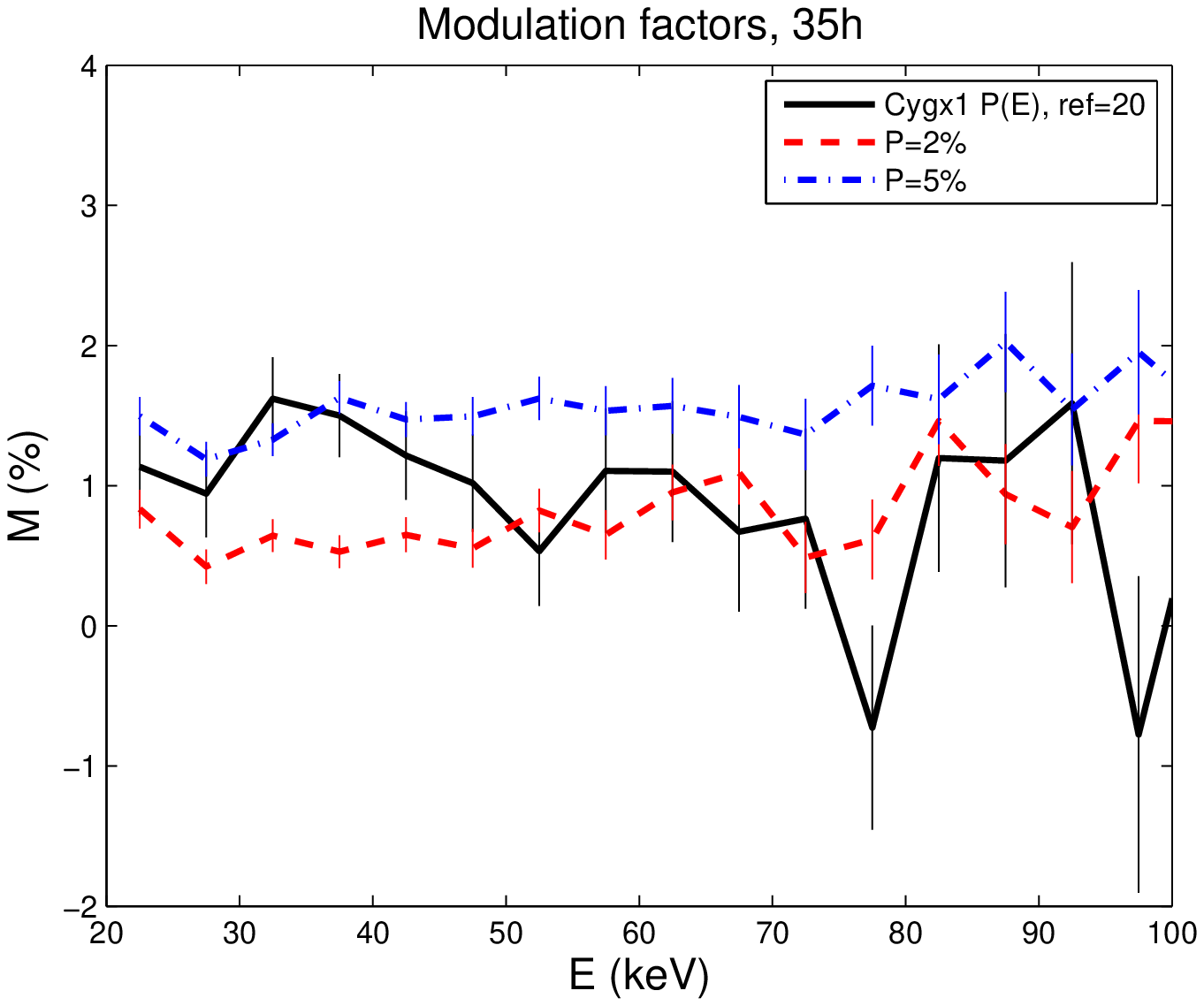}
  \includegraphics[width=7cm]{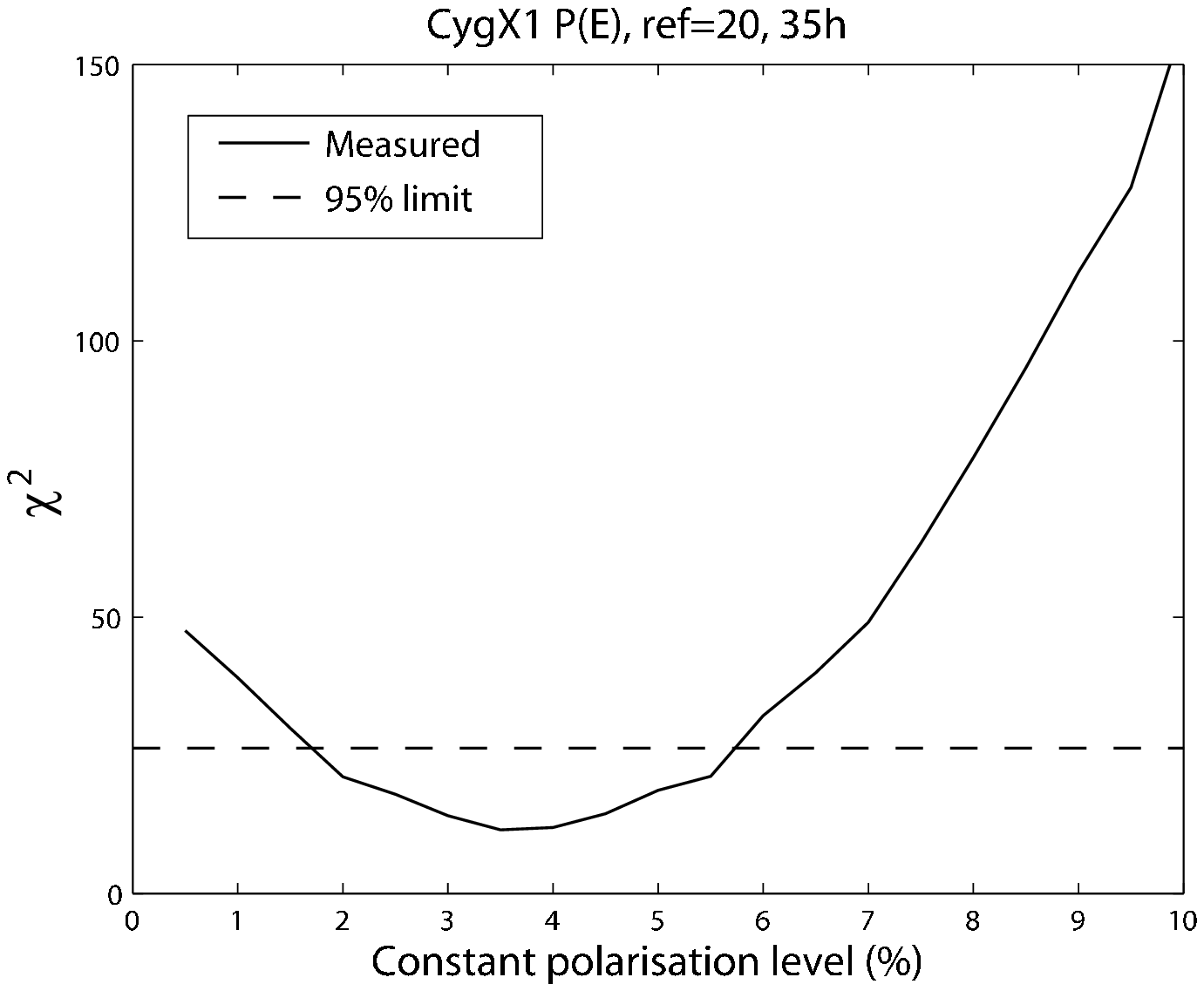}
  \caption{Same as Fig.~\ref{cyg100results}, but with the reflection component assumed to be 20\% polarized. \textit{Left panel:} Expected modulation factor $M(E)$, together with models of constant polarization at 2\% and 5\%. \textit{Right panel:} $\chi^2$ values when comparing $M(E)$ for the simulation with different degrees of constant polarization. The dashed line marks the limit for 95\% significance in rejecting constant polarization. Constant polarization cannot be rejected with high significance.}
  \label{cyg20results}
\end{figure*}

Table~\ref{tab:cyg result} summarises the results for the simulations of Cygnus~X-1. For a 100\% polarized reflection component, the energy dependence is detected both after 6 hours of observation and after 35 hours. Figure~\ref{cyg100results} shows the expected modulation factors and $\chi^2$ values for a 35h observation. In the case of 20\% polarization of the reflection component, shown in Fig.~\ref{cyg20results}, constant polarization cannot be ruled out at any higher significance level.

\begin{table}[bt]
  \begin{tabular*}{\columnwidth}{@{\extracolsep{\fill}}rll|ll}
  &  \multicolumn{2}{c}{Reflection 100\%} &
\multicolumn{2}{l}{Reflection 20\%} \\
    \hline
    Obs. time \vline & 6h & 35h & 6h & 35h \\
    Significance \vline & 99.4\% & $>$99.99\% & 19.1\% & 51.2\% \\
\hline
  \end{tabular*}
  \caption{The significance in rejecting constant polarization models, shown for different reflection polarization strengths and observation times in the case of Cygnus X-1.}
  \label{tab:cyg result}
\end{table}

\subsection{Magnetic NS}

The results for the simulations of the accreting neutron star are summarised in Table \ref{tab:ns result}. For a 100\,mCrab source, the background will start to become significant already in the higher end of the PoGOLite energy range. To be conservative, only $M(E)$ data points up to 60 keV were used for the $\chi^2$ analysis in order to limit the errors due to uncertainties in the background flux. This is equivalent to 8 degrees of freedom, demanding $\chi^2>15.5$ for 95\% certainty in rejecting constant polarization. 

\begin{table}[tb]
  \begin{tabular*}{\columnwidth}{@{\extracolsep{\fill}}r|lllll}
    \multicolumn{6}{c}{Gaussian peaks} \\
    \hline
    $\Pi_\textrm{max}$ (\%) & 10 & 20 & 30 & 40 & 50 \\
    Significance & 24.3\% & 24.3\% & 35.3\% & 79.9\% & 99.7\% \\
    \hline
    \multicolumn{6}{c}{Positive steps} \\
    \hline
    $\Pi_\textrm{max}$ (\%) & 10 & 20 & 30 & 40 & 50 \\
    Significance & 46.3\% & 70.0\% & 96.8\% & 91.8\% & 83.5\% \\
    \hline
    \multicolumn{6}{c}{Negative steps} \\
    \hline
    $\Pi_\textrm{max}$ (\%) & 10 & 20 & 30 & 40 & 50 \\
    Significance & 51.0\% & 99.3\% & 95.8\% & 99.6\% & $>$99.99\% \\
    \hline
  \end{tabular*}
  \caption{The significance in rejecting constant polarization models, shown for different polarization shapes and maxima in the case of a 100\,mCrab neutron star.}
  \label{tab:ns result}
\end{table}

\subsubsection{Gaussian peaks}
\begin{figure*}
  \centering
  \includegraphics[width=8cm]{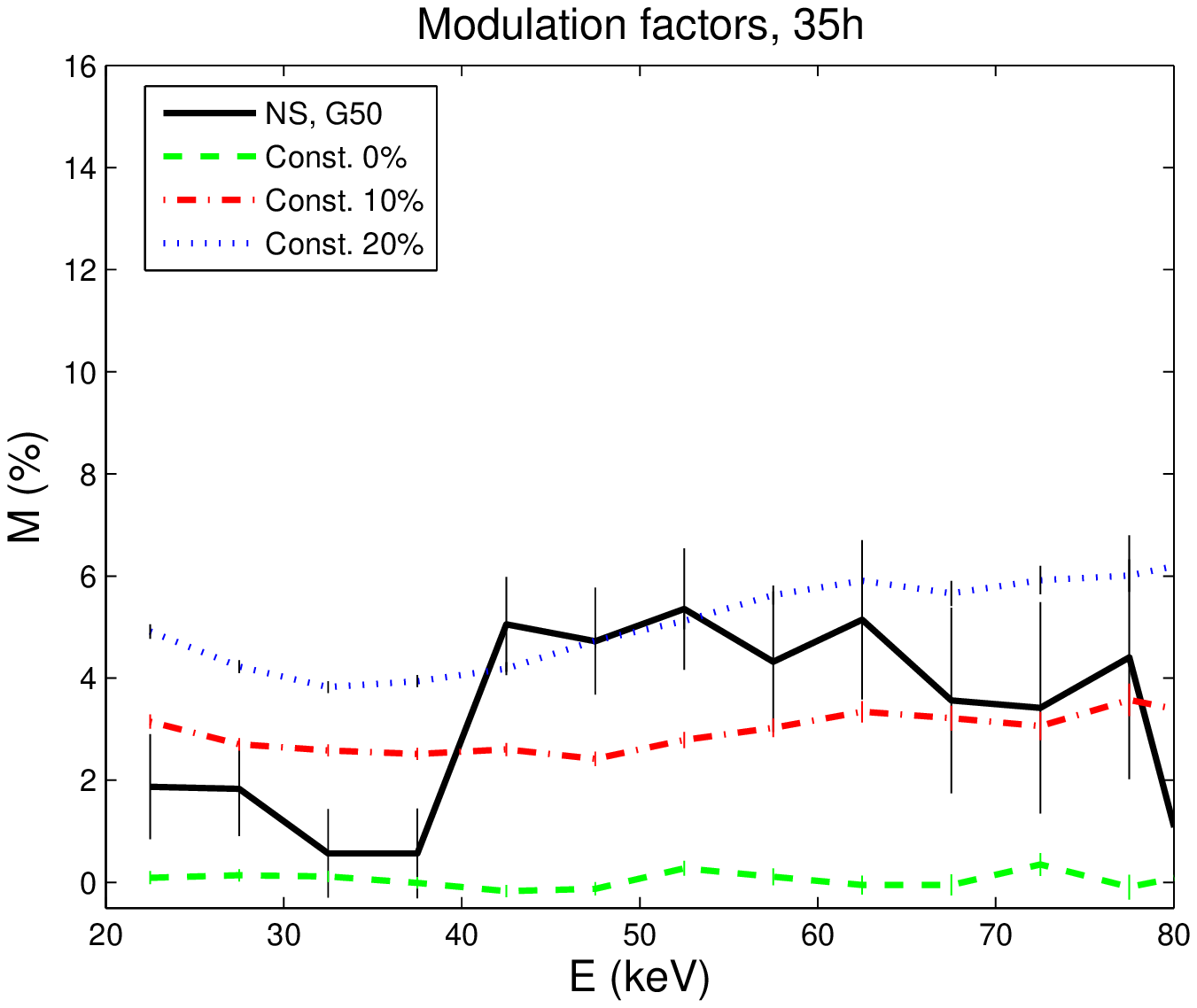}
  \includegraphics[width=7cm]{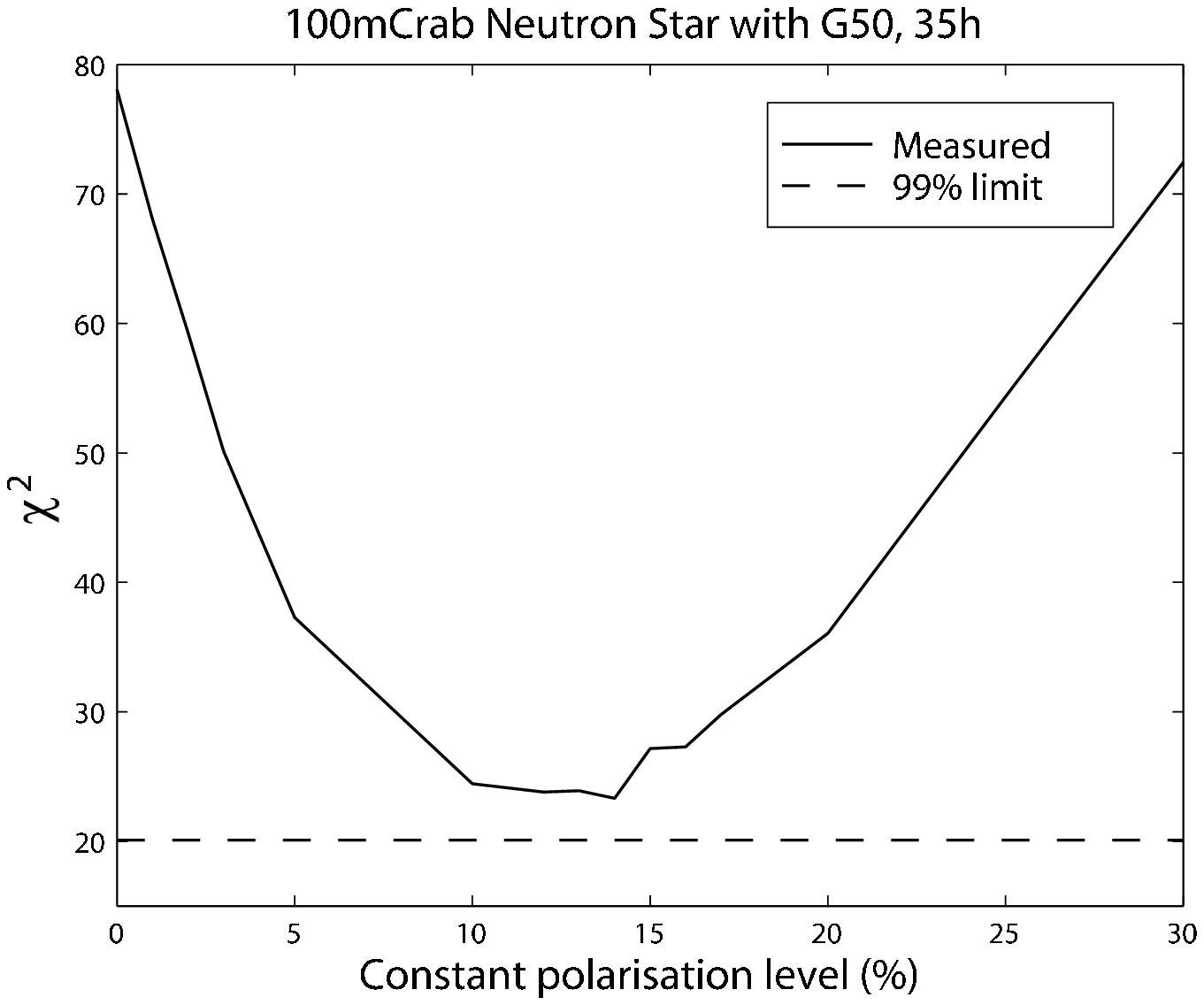}
  \caption{Results from simulations of a 35h observation of an accreting magnetic neutron star, assuming a Gaussian polarization curve. \textit{Left panel:} Expected modulation factor $M(E)$, together with models of constant polarization at 0\%, 10\% and 20\%. The modulation is fitted in intervals of 5 keV. \textit{Right panel:} $\chi^2$ values when comparing $M(E)$ for the simulation with different degrees of constant polarization. The dashed line marks the limit for 95\% significance in rejecting constant polarization. Constant polarization is rejected with high significance.}
  \label{nsgresults}
\end{figure*}

In Fig.~\ref{nsgresults} we see $M(E)$ curves for $G_{50}$ and a few constant polarizations (left panel), clearly illustrating their different characteristics. The resulting $\chi^2$ curve (right panel) confirms that these models are significantly (99\% level) different. However, this was not the case for any lower value of $\Pi_{\rm max}$.

\subsubsection{Positive steps}
\begin{figure*}
  \centering
  \includegraphics[width=8cm]{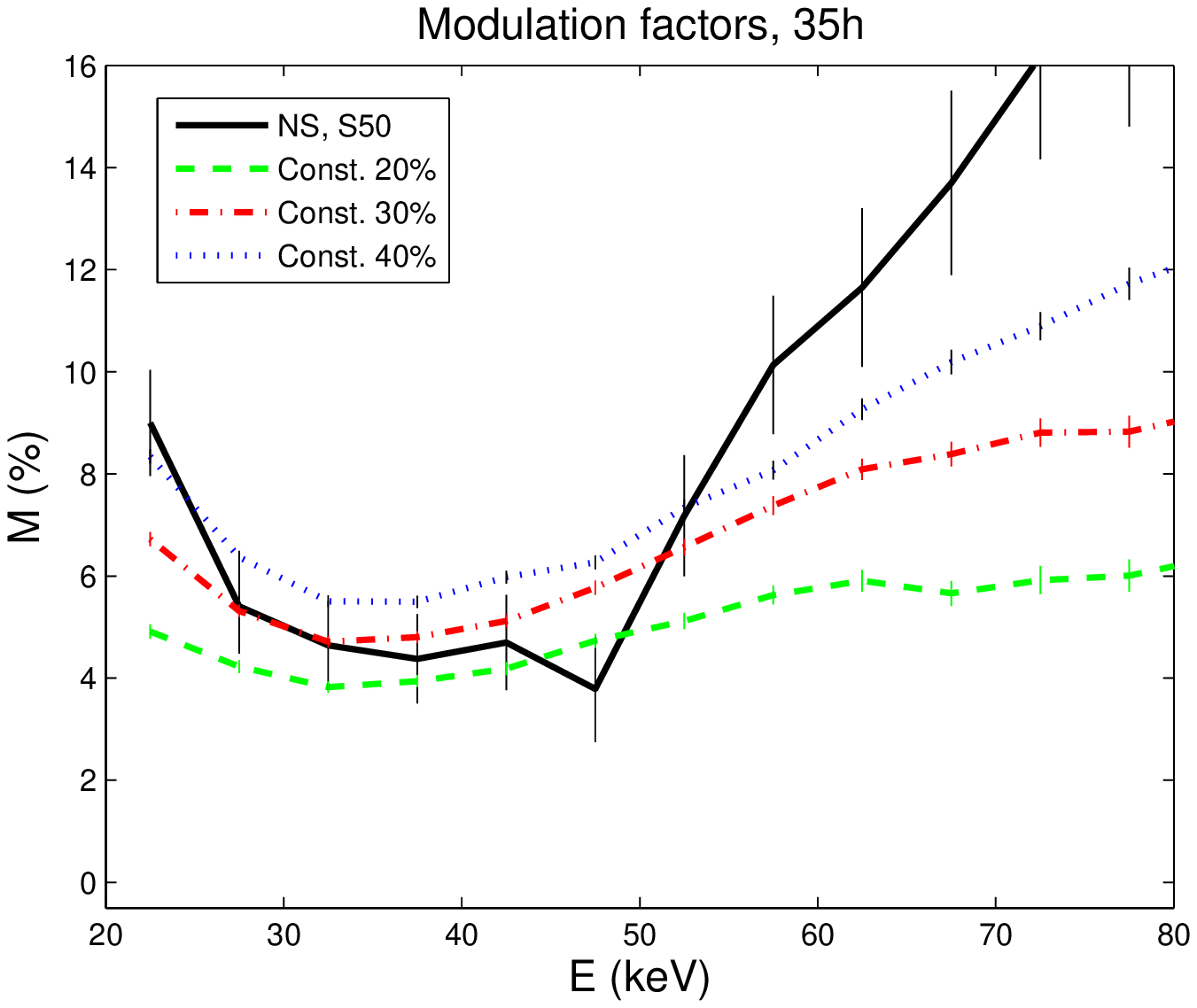}
  \includegraphics[width=7cm]{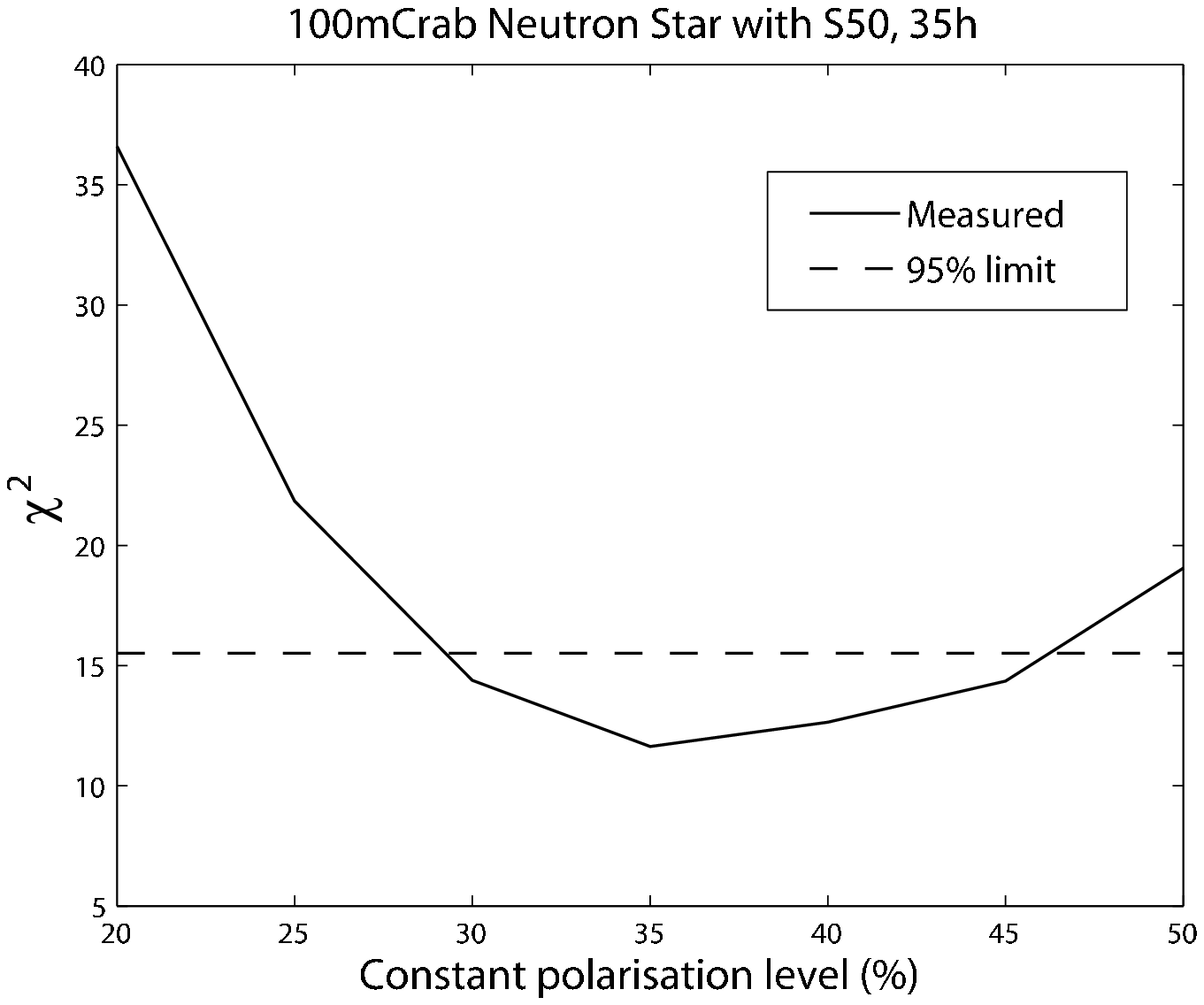}
  \caption{Results from simulations of a 35h observation of an accreting magnetic neutron star, assuming a positive step with $\Pi_{\rm max}=50\%$. \textit{Left panel:} Expected modulation factor $M(E)$, together with models of constant polarization at 20\%, 30\% and 40\%. The modulation is fitted in intervals of 5 keV. \textit{Right panel:} $\chi^2$ values when comparing $M(E)$ for the simulation with different degrees of constant polarization. The dashed line marks the limit for 95\% significance in rejecting constant polarization. Constant polarization cannot be rejected with high significance.}
  \label{ns50uresults}
\end{figure*}

Figure~\ref{ns50uresults} shows the result for $S_{50}$ when compared to cases of constant polarization. In this case, the difference in characteristics between the constant models and the energy dependent model is not large enough, and constant polarization cannot be ruled out. The only value of $\Pi_{\rm max}$ yielding a significant (95\% level) difference was $\Pi_{\rm max}=30\%$. What is interesting to note is that a higher $\Pi_{\rm max}$, which implies a higher total polarization, does not necessarily make the energy dependence easier to measure.

\subsubsection{Negative steps}
\begin{figure*}
  \centering
  \includegraphics[width=8cm]{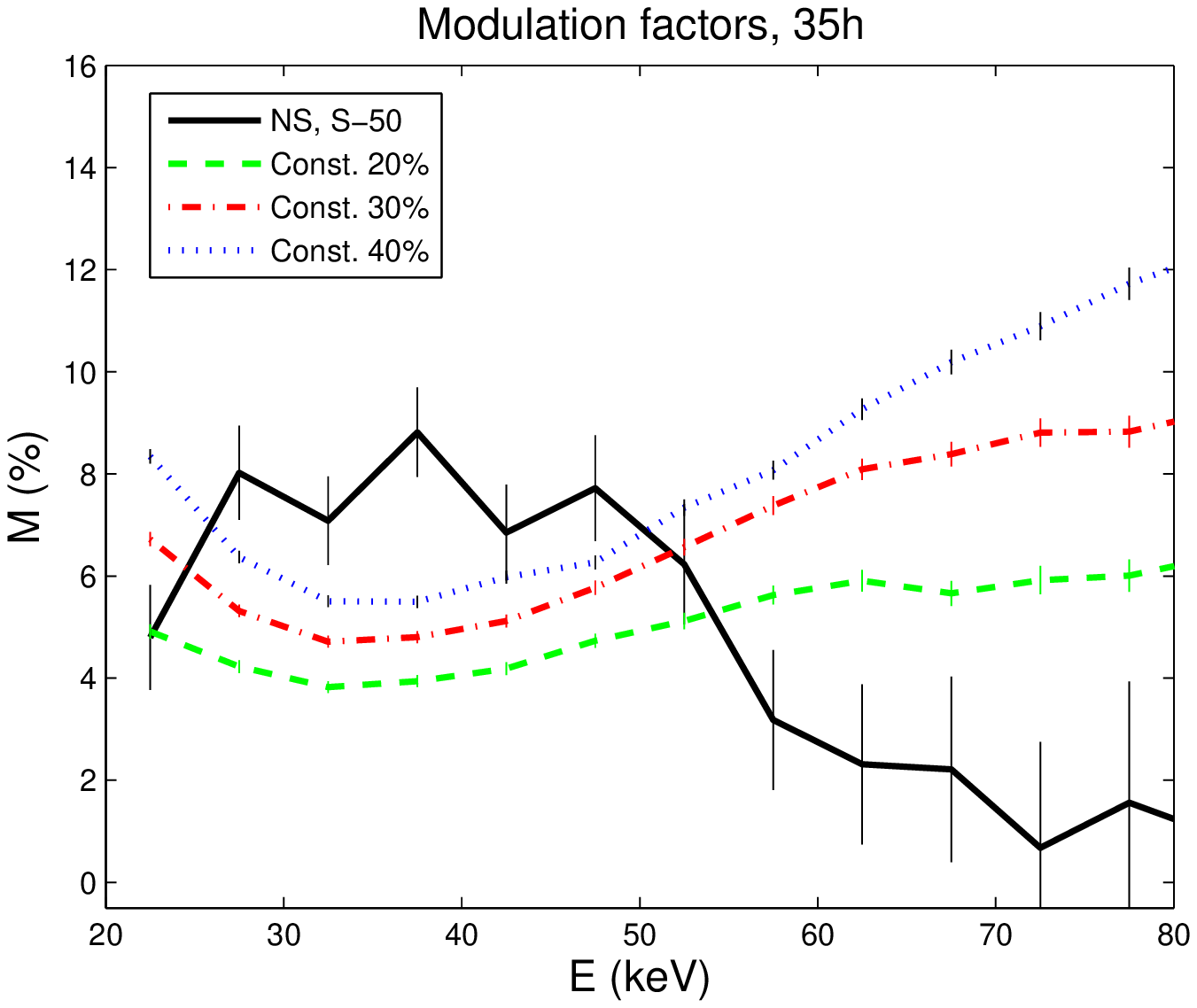}
  \includegraphics[width=7cm]{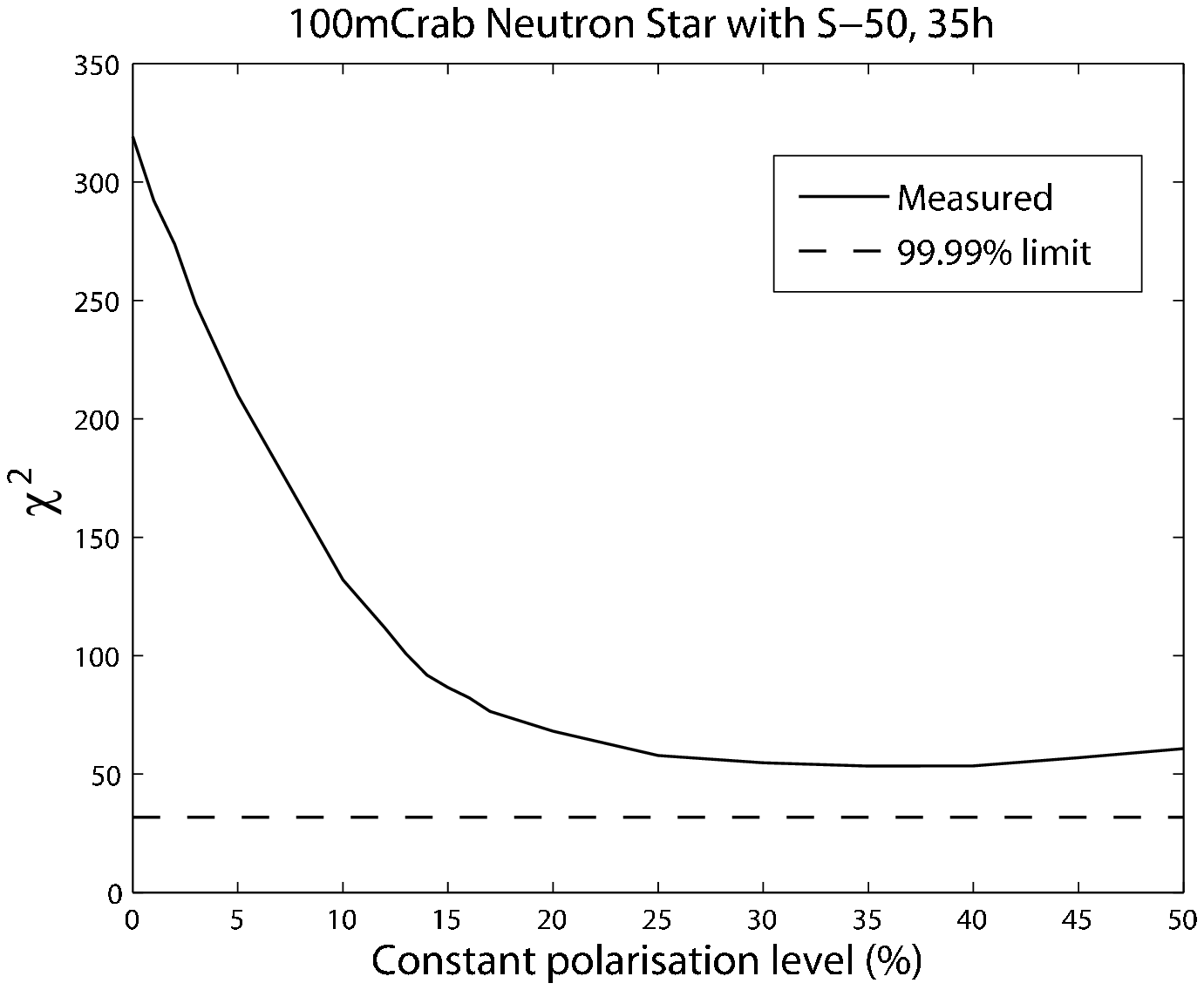}
  \caption{Same as in Fig.~\ref{ns50uresults}, but using the negative step. In this case, constant polarization is rejected with high significance.}
  \label{ns50dresults}
\end{figure*}

The result of the $S_{-50}$ model is shown in Fig.~\ref{ns50dresults}, where the $\chi^2$ plot (right panel) allows us to reject constant polarization with a certainty much higher than 99.99\%. Other models yielding significant detections were $S_{-20}$ (99\%), $S_{-30}$ (95\%), and $S_{-40}$ (99\%). One reason these models are so easily measurable stems from the fact that, due to the shape of the neutron star spectrum, data and statistics are much poorer at higher energies. If the polarized part only lies at high energies, the total polarization will be much lower or even undetectable.

\section{Discussion}
\label{discussion}
Although no measurements have been made of polarization in the energy range covered by PoGOLite, some theoretical models predict changes in polarization with energy. The results of the simulations clearly show that it is possible for the PoGOLite instrument to detect the energy dependence of polarization for several of the investigated cases. The highest significance is found for Cygnus~X-1 assuming a fully polarized reflection component and a neutron star in the case of a negative step; in the remaining cases, constant polarization can not be rejected. Our results therefore show that PoGOLite has the potential to discriminate among these models.

As described above, in Cygnus~X-1 the hard X-rays are believed to originate from Comptonization of soft seed photons in a predominantly thermal electron distribution. Although this process involves Compton scattering -- which could introduce a net polarization -- multiple scatterings are required, making our assumption of the direct component being unpolarized a reasonable one. The degree of polarization of the reflected component is however more uncertain. Our idealized case of 100\% polarization is certainly an overestimation. Calculations \cite{matt93} show that the degree of polarization in the reflected component varies with inclination, with a maximum of $\sim 30\%$ expected at high inclination. The inclination of the Cygnus~X-1 system is not well known, but estimates put it at $30^\circ$--$50^\circ$ \cite{mil02}, making our assumption of 20\% polarization in the reflected component reasonable. The relative size of the reflected component compared to the direct emission is in turn dependent on both inclination, where the dependence is the opposite one, and system geometry. We note that in other sources the reflected component may be much stronger, or even dominate the radiation spectrum (e.g., Cygnus~X-3, \cite{hja07}).  

Another issue which may complicate measurements of energy dependent polarization is the behaviour of the angle of polarization. In our simulations, we have implicitly assumed that the angle does not vary with energy. However, for both black hole and neutron star systems, this assumption may be an oversimplification. It is certainly true that emission originating from the region close to a black hole will be affected by the strong gravity, affecting the polarization angle \cite{conn80}. It is not clear how large this effect would be on the reflected component in, for instance, Cygnus~X-1, but results from both temporal and spectral analysis show that the accretion disc -- assumed to be the reflector -- is presumably truncated at a large distance ($R_{\rm in} \gtrsim 30 R_{g}$, \cite{axe06,gie97}) from the black hole in the hard state. We therefore do not expect this effect to alter the outcome of our simulations. 

Our results from simulating Cygnus~X-1 indicate that long observations with PoGOLite are required to search for energy dependence of polarization. The first flights of the instrument will likely be shorter flights, covering several targets. While these observations should be long enough to detect polarization down to the level of a few per cent, we do not expect to detect any changes in polarization with energy. However, long duration flights spanning several days are also planned, and such flights would provide the observation time needed to search for variations of polarization degree with energy.

A point to note from the neutron star simulations is the result that it is easier to rule out constant polarization in the case of a negative step than for a positive step. As noted in Sect.~\ref{results}, the energy response of the instrument is such that the energy of the incoming photon cannot be uniquely determined. The result will be a redistribution of energy, with a possibility for higher energy photons to be detected with lower energies. The reverse is however not true -- a low energy photon will not be detected as having a higher energy. In the case of a positive step, some polarized high energy photons will be detected at lower energies. This will give a false polarization signal at lower energies, and act to ``smooth'' the detected energy dependance of polarization. For a negative step, the polarization is introduced at lower energies and will not ``spread'' to higher energies. 
The low energy polarization will be somewhat diluted by
redistribution of high energy photons but the polarization
contrast will still be higher than in the positive case.

By excluding energies above 60 keV in the ${\chi}^2$ analysis
of accreting X-ray pulsars we have been fairly conservative
in our estimate of significances. The restriction of the
energy range was motivated by a potentially high sensitivity to
systematic errors in the background level at high energies.
In this analysis however, we have not taken advantage of
the fact that these sources are pulsating. By analysing
the polarization of the pulsed flux, rather than the total
flux, it should be possible to include all points up to
100 keV and thereby increase the sensitivity.
On the other hand the polarization direction will
probably change over the pulsation period which will have
the opposite effect of reducing the sensitivity. How
important this effect is depends on the precise source
geometry, radiation beaming and our viewing angle.

As PoGOLite's field-of-view is rather large (\mbox{$\sim5\;{\rm deg}^2$}),
the pointing errors with respect to the axis of rotation must be small
to avoid introducing systematic errors in the polarization measurements.
The attitude control system used for PoGOLite will assure accurate 
pointing to within a few arcminutes, keeping the systematic error below 
1\% \cite{kam07}. Although this figure refers to the whole energy band,
we do not expect any such effect to change the results presented here.
A comprehensive study of systematic effects is beyond the scope of
this paper, but will be crucial once PoGOLite is in operation. 

The performance of the PoGOLite instrument has been extensively evaluated, 
both with laboratory-based tests \cite{mozsi}, accelerator-based tests 
\cite{kan07}, and simulations \cite{olle}. These tests show that it
will be able to detect low ($\sim 10\%$) degrees of polarization even
for 100 mCrab sources. What has not previously been tested is its
sensitivity to a polarization degree that varies with energy.
Despite the relatively modest inherent energy resolution, our results 
show that PoGOLite has the capacity
to detect changes in polarization degree with energy. The simulations
show that significant results can be obtained in a 35h observation,
attainable in the long duration flights already planned for PoGOLite.
We stress that the design is not optimized for such detections, and
future instruments will in all likelihood develop this technique
further.

\section{Conclusions}
\label{conclusions}

The Compton technique applied to an array of plastic scintillators is an 
effective method to measure broad energy band polarization, which is 
demonstrated with the proposed
PoGOLite mission, using Geant4 simulations. In particular, energy
dependence can be detected. However, in our model of polarization from
X-ray binaries, we require the reflection to contribute a large 
fraction of the observed flux and/or have high degree of polarization for
energy dependence to be detected. Similarly, for accreting magnetic neutron 
stars, sharp energy variations in the polarization are needed for a clear
detection. This is made easier if the lower energies contain most of
the polarization.

\section*{Acknowledgments}

The authors gratefully acknowledge support from the Knut and Alice Wallenberg Foundation, the Swedish National Space Board, the Swedish Research Council, the Kavli Institute for Particle Astrophysics and Cosmology (KIPAC) at Stanford
University through an Enterprise Fund, and the Ministry of Education, Science, Sports and
Culture (Japan) Grant-in-Aid in Science No.18340052. J.K. and N.K. 
acknowledge support by JSPS.KAKENHI (16340055). J.K. was also supported by a
grant for the international mission research, which was provided by the 
Institute for Space and Astronautical Science (ISAS/JAXA). T.M. acknowledges 
support by Grants-in-Aid for Young Scientists (B) from Japan Society for 
the Promotion of Science (No. 18740154).

\end{document}